\documentclass{PoS}

\title{Constraints on Decaying Dark Matter}

\ShortTitle{Constraints on Decaying Dark Matter}

\author{\speaker{Annika H. G. Peter}\thanks{This work was supported by the Gordon and Betty Moore Foundation and by  DOE Grant No. DE-FG03-92-ER40701.}\\
        California Institute of Technology MC 249-17, Pasadena, CA 91125, USA\\
        Department of Physics \& Astronomy, University of California, Irvine, CA 92697, USA\\
        E-mail: \email{apeter@astro.caltech.edu}}

\author{Christopher E. Moody\\
        Department of Physics, University of California, Santa Cruz, California 95064, USA\\
        E-mail: \email{cemoody@ucsc.edu}}

\author{Andrew J. Benson\\
        California Institute of Technology MC 350-17, Pasadena, CA 91125\\
        E-mail: \email{abenson@its.caltech.edu}}

\author{Marc Kamionkowski\\
        California Institute of Technology MC 350-17, Pasadena, CA 91125\\
        E-mail: \email{kamion@tapir.caltech.edu}}

\abstract{We explore a dark-matter model in which there are two dark-matter species nearly degenerate in mass, with $\epsilon = \Delta M/M 
\ll 1$. The heavier particle undergoes two-body decay with a half-life $\tau$, to the lighter dark-matter particle and a noninteracting massless particle. Unlike previous work on decaying dark matter, we explore the regime $\tau > 100$ Myr and non-relativistic kick speeds $v_k / c = \epsilon$. Using a set of N-body simulations of isolated dark-matter halos, we show how halos change as a function of $\tau$ and $v_k$. We find that $\tau < 40$ Gyr is ruled out for $v_k \gtrsim 20 \hbox{ km s}^{-1}$ ($\epsilon \gtrsim 10^{-4}$) when we compare the simulations to observations of dwarf-galaxy- to cluster-mass dark matter halos.  We highlight which set of observations should provide better future constraints for decays and other types of dark-matter physics.}

\FullConference{Identification of Dark Matter 2010-IDM2010\\
		July 26-30, 2010\\
		Montpellier France}

\begin{document}

\newcommand{\Rvir}{$R_{\mathrm{vir}}$}
\newcommand{\Rvire}{R_{\mathrm{vir}}}
\newcommand{\Mvir}{$M_{\mathrm{vir}}$}
\newcommand{\Mvire}{M_{\mathrm{vir}}}
\newcommand{\tdyn}{$t_{\mathrm{dyn}}$}
\newcommand{\tdyne}{t_{\mathrm{dyn}}}
\newcommand{\vk}{$v_{\mathrm{k}}$}
\newcommand{\vke}{v_{\mathrm{k}}}
\newcommand{\vvir}{$v_{\mathrm{vir}}$}
\newcommand{\vvire}{v_{\mathrm{vir}}}
\newcommand{\rhose}{\rho_{\mathrm{s}}}
\newcommand{\rse}{r_{\mathrm{s}}}
\newcommand{\Deltave}{\Delta_{\mathrm{v}}}
\newcommand{\rhoue}{\rho_{\mathrm{u}}}
\newcommand{\rhe}{r_{\mathrm{h}}}
\newcommand{\zre}{z_{\mathrm{re}}}
\newcommand{\vmax}{$v_{\mathrm{max}}$}
\newcommand{\vmaxe}{v_{\mathrm{max}}}
\newcommand{\Modot}{\mathrm{M}_\odot}

\section{Introduction}
Most of the gravitationally attractive mass-energy of the Universe consists of dark matter, but the physics of dark matter is poorly characterized \cite{komatsu2010}.  The most popular dark-matter candidate class, the weakly-interacting massive particle (WIMP), consists of particles that are heavy ($m\gtrsim 100\hbox{ GeV}$), self-annihilating, stable, and have low interaction probabilities with other particles \cite{steigman1985}.  However, there are a number of other well-motivated candidates with different phenomenology \cite{feng2003,feng2008}.  Dark matter may have strong elastic self-interactions, it may not self-annihilate, or may be unstable \cite{spergel2000,cohen2010,kaplinghat2005}.  The identification of dark matter depends on uncovering its phenomenology.

In this proceeding, we discuss constraints on the stability of dark matter.  In particular, we consider a dark-matter model that consists of two nearly degenerate WIMP species, $X$ and $Y$, in which $X$ decays to $Y$ and a (nearly) massless particle on cosmological time scales.  This is one of the simplest extensions to the WIMP model.  The mass difference $\epsilon = \Delta M/M$ is related to the recoil speed of the daughter $Y$ particle $\vke/c = \epsilon$ if $\epsilon \ll 1$.  There are thus three parameters for this model: $M_X$, the mass of the $X$ particle; $\vke$ (or $\epsilon$), the recoil speed of the daughter $Y$; and $\tau$, the half life.  

While constraints on this model are stringent if the (nearly) massless particle in this model is a gamma ray or decays to light leptons \cite{bell2010}, there are few constraints if the (nearly) massless particle has negligible interactions or is a (or decays to) neutrino.  To constrain a generic nearly-degenerate decay model, we focus on the structure and evolution of dark-matter halos.  Decay can affect halos in two ways: daughter particles can escape, causing mass loss; and the dark-matter density decreases within the halo in response to mass loss or, in the case the daughter particle remains bound, to the difference in kinetic energy between the parent and daughter particles.  In practice, this means that the value of $M_X$ is irrelevant as long as it is similar to WIMP-range masses.

We use simulations to characterize the evolution of dark-matter halos in response to decays, as a function of \vk~and $\tau$; these are described briefly in Sec. \ref{sec:sim} and in more detail in Refs. \cite{peter2010b,peter2010c}.  In Sec. \ref{sec:results} we use the simulations as a map the decay parameters to observables in order to use the properties of dark-matter halos inferred from observations of galaxies, groups, and clusters to constrain the decay parameter space.  We discuss our key findings in Sec. \ref{sec:discussion}.

\section{Simulations}\label{sec:sim}
We simulate isolated equilibrium dark-matter halos using the
parallel $N$-body code GADGET-2, modified by us to handle decays \cite{springel2001a,peter2010b}.  Since we consider $\tau$ to be long, we assume that halos collapse before there is significant decay.  Thus, the initial conditions for the halos will be well-described by the expected properties of cold-dark-matter (CDM) halos.  Our halos are spherically symmetric, and the density $\rho(r)$ as a function of galactocentric radius $r$ is taken to be the
Navarro-Frenk-White (NFW) form \cite{navarro1996},
\begin{eqnarray}\label{eq:rho}
	\rho(r) = \frac{\rho_\mathrm{s}}{\displaystyle
        \frac{r}{r_\mathrm{s}}\left(1+\frac{r}{r_\mathrm{s}}\right)^{2}},
\end{eqnarray}
where $\rho_\mathrm{s}$ is the scale density, and $r_\mathrm{s}$
is the scale radius, such that the halo concentration is
\begin{eqnarray}\label{eq:c}
	c = \frac{\Rvire}{\rse},
\end{eqnarray}
where \Rvir~is the virial radius of the halo.  We define virial quantities with respect to the spherical top-hat overdensity, as described by Ref. \cite{bryan1998}.

The initial velocity distribution must be chosen carefully in order for the halo to be in dynamical equilibrium in the absence of decays.  We take the velocity ellipsoid to be isotropic, a reasonable
first approximation for dark-matter halos \cite{diemand2009}.  The procedure to initialize the velocities is described in more detail in Ref. \cite{peter2010b}.

We begin every simulation with identical initial conditions, with $10^6$ particles in the virial radius, varying only the concentration $c$ of the NFW profile, particle
lifetimes, and kick velocities.  We simulate virial mass $\Mvire =
10^{12}M_\odot$ halos only, but we can extrapolate our results
to other halo masses by considering the structural changes to
the halos as a function of the decay parameters with respect to the halo virial speed and dynamical time.  We perform a set of 100
simulations, with $\vke = 10,$ 50, 100, 200, and 500 km s$^{-1}$
and $\tau = 0.1,$ 1, 10, 50, and 100 Gyr, and with $c=5,10,20$ and $30$.  Since the virial speed of the halo is $\vvire \equiv
\sqrt{G\Mvire/\Rvire} \approx 130\hbox{ km s}^{-1}$ using a
spherical top-hat overdensity definition of virial parameters,
the kick speeds were chosen to bracket the virial value.  The simulations are discussed in depth in Refs. \cite{peter2010b,peter2010c}.

\section{Results}\label{sec:results}
In this section, we show what kind of constraints one can set on the decay parameter space given the observations described below.  Those interested in the details of the evolution of the dark-matter halo density and velocity-dispersion profiles as a function of decay parameters should consult Sec. III of Ref. \cite{peter2010b}.  

\subsection{Cluster Mass Function}
The comoving galaxy-cluster mass function $dn/dM$ is a sharply falling function of halo mass at all redshifts, and is a probe of both the Hubble expansion (via the conversion of redshifts to comoving volumes) and the growth function of structure \cite{albrecht2006}.  In the absence of decays or modifications to gravity, the cosmological parameters inferred from probes of homogeneous cosmology (e.g., the Hubble expansion) should agree with those inferred from probes of the growth of structure, and those parameter values should be consistent across cosmic time.  However, decays change the growth of structure, with mass loss from halos becoming increasingly important with time.  If the decays are non-relativistic, then the Hubble expansion should not deviate from what would be predicted from the cosmic microwave background, but the growth function would be different.

\begin{figure}
  \begin{center}
    \includegraphics[width=0.47\textwidth]{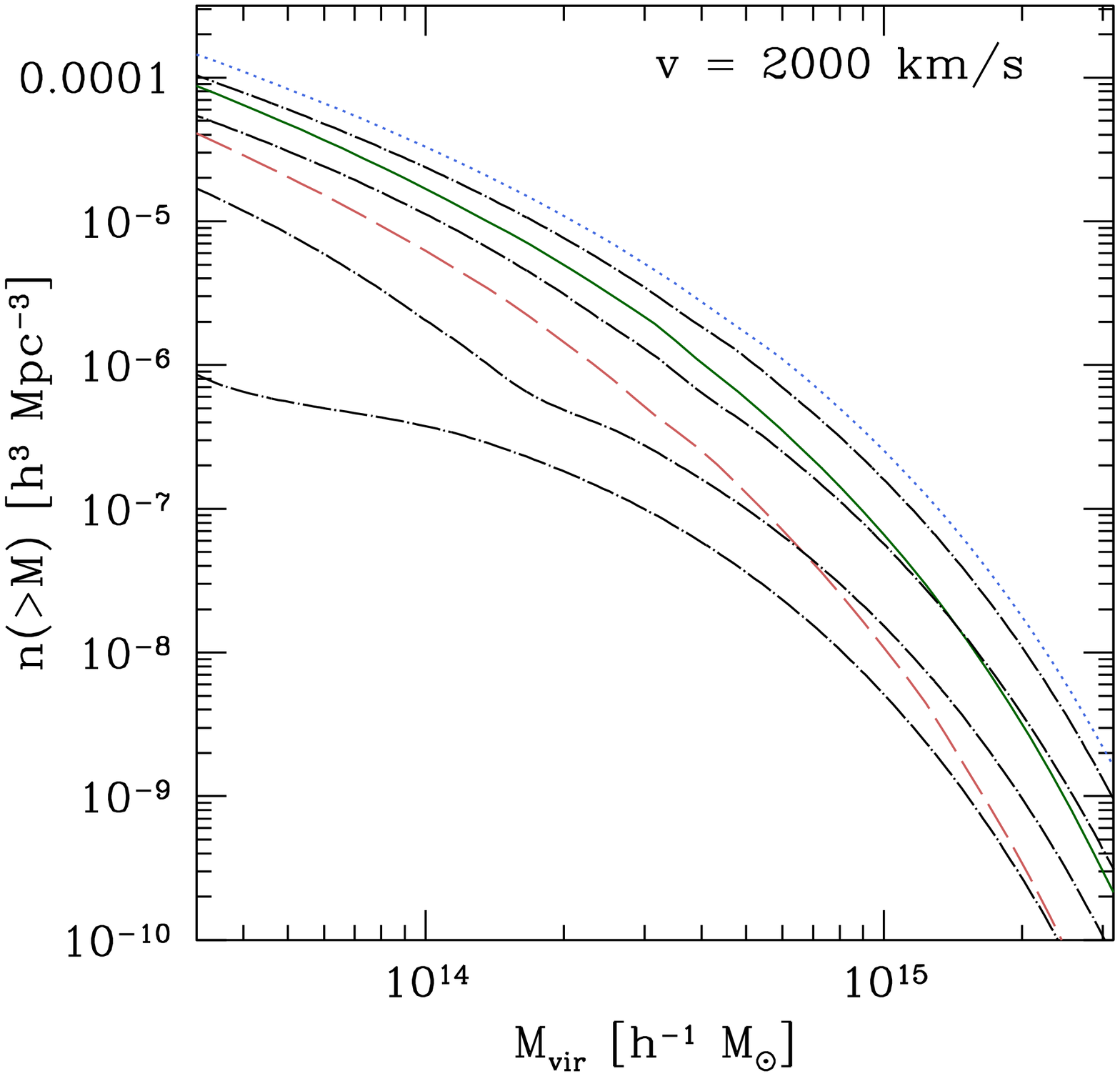}  \includegraphics[width=0.47\textwidth]{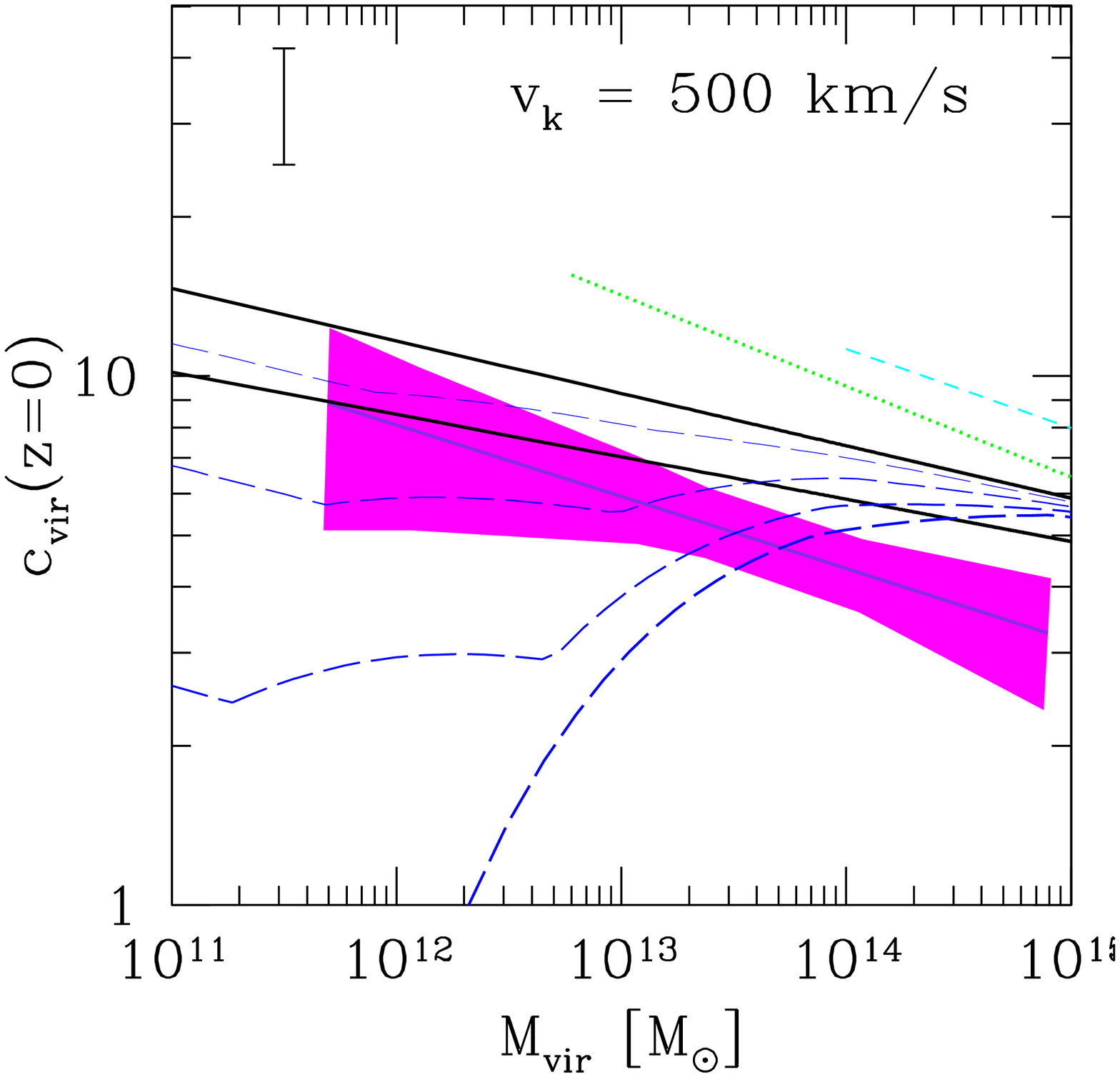}
  \end{center}
  \caption{\label{fig:lss}Examples of constraints on decays from (\emph{left}) the $z=0$ cluster mass function and (\emph{right}) the $z=0$ mass-concentration relation.  See text for details.}
\end{figure}

To constrain decays using the cluster mass function, we take the highest-normalization $z=0$ cluster mass function allowed by high-$z$ probes of cosmological parameters (e.g., the cosmic microwave background \cite{komatsu2010}) assuming a $\Lambda$CDM cosmology, and map the CDM halo masses to halo masses after a Hubble time of decay using our simulations.  In order for a point in decay parameter space to be allowed, the $z=0$ cluster mass function from the decay simulations must lie above the minimum cluster mass function allowed by X-ray and optical observations of low-redshift clusters \cite{vikhlinin2009}.  We show an example of constraints in the left-hand plot of Fig. \ref{fig:lss}, in which the comoving number of clusters above mass \Mvir~is plotted against \Mvir.  The colored lines indicate (top to bottom) the highest cluster mass function consistent with \emph{WMAP} seven-year data, the mean cluster mass function predicted from \emph{WMAP}, and the 1$-\sigma$ lowest mass function consistent with $z=0$ observations of clusters \cite{vikhlinin2009}.  The dot-dashed lines indicate cluster mass functions for $\vke = 2000\hbox{ km s}^{-1}$ and (top to bottom) $\tau = 100, 40, 20,$ and 5 Gyr.  In general, $\tau \gtrsim 30$ Gyr for $\vke \gtrsim 1000\hbox{ km s}^{-1}$ ($\epsilon \gtrsim 10^{-2}$) in order to be consistent with the observed cluster mass function.

One can also use the redshift-dependent cluster mass function to probe decays (see, e.g., \cite{peter2010a}).  The strength of this approach is that the deviations of the cluster mass function from a $\Lambda$CDM model chosen to match $z=0$ observations should increase with look-back time.  The challenge of this approach is that the redshift-dependent cluster mass function will also probe non-$\Lambda$ models for the accelerated expansion, and there may be degeneracies between decay and modified gravity in the redshift evolution of the cluster mass function.

\subsection{Mass-Concentration Relation}
The mass distribution within dark-matter halos depends on the formation time; small halos that form early when the Universe is more dense have a higher central-density-to-virial-density ratio than larger halos that formed later, causing a mass-concentration relation \cite{bullock2001}.  High-normalization, high-$\Omega_\mathrm{m}$ cosmologies tend to produce higher concentrations for fixed mass than for low-nor\-mal\-ization, low-$\Omega_\mathrm{m}$ cosmologies because the typical formation time of halos is earlier.  Decays lower the concentrations of dark-matter halos due to the negative heat capacity of self-gravitating systems, so the observed mass-concentration relation should provide good constraints on decay.

We show how decays affect the mass-concentration relation in the right-hand plot of Fig. \ref{fig:lss}.  On this plot, the best-fit NFW concentration (Eq. \ref{eq:c}) is plotted against the virial mass of halos at $z=0$, with halos spanning the range of those hosting $L^*$ galaxies to massive clusters.  The upper and lower solid lines show the predicted mass-concentration relation for \emph{WMAP} one- and three-year cosmologies, respectively, the former of which is approxately the 2$-\sigma$ upper limit from \emph{WMAP} seven-year data \cite{maccio2008}.  The short-dashed and dotted lines show estimates of the mass-concentration relation from strong-lensing and X-ray observations of clusters \cite{buote2007}.  The selection function of the observations is poorly understood but is generally thought to prefer high-concentration, early-forming relaxed clusters.  The magenta-shaded region represents the 1$-\sigma$ uncertainty in the central mass-concentration relation from weak lensing assuming a $\Lambda$CDM cosmology \cite{mandelbaum2008}.  The error bar in the top-left corner of the plot indicates the scatter in the simulated mass-concentration relation and the typical 1$-\sigma$ uncertainty in the fit for the mass-concentration relation from observations.

To find the most conservative constraints on decay, we assume a high-normalization, high-$\Omega_\mathrm{m}$ Universe consistent at the 2$-\sigma$ level with the \emph{WMAP} seven-year analysis.  We consider a point in decay parameter space to be ruled out if its corresponding mass-concentration relation lies below the magenta-shaded region of Fig. \ref{fig:lss}, since weak lensing has a better-understood selection function than strong lensing or X-ray clusters and groups.  In the right-hand panel of Fig. \ref{fig:lss}, we consider $\vke = 500\hbox{ km s}^{-1}$, with the black dashed lines corresponding to (top to bottom) $\tau = 100, 40, 20,$ and 5 Gyr.  In general, we find $\tau \lesssim 30$ Gyr to be excluded for $\vke \gtrsim 200\hbox{ km s}^{-1}$.  This is more exclusive than for the cluster mass function because weak lensing probes much smaller dark-matter halos, and decays have the greatest effect on halo structure if $\vke \gtrsim \vvire$.

\subsection{Local Group Dwarf Galaxies}
The Local Group has traditionally been a source of inspiration for dark-matter theories, especially since by 1999, simulations with $\Lambda$CDM cosmologies indicated that there ought to be far more subhalos in the Milky Way halo than there were satellite galaxies observed \cite{moore1999}.  However, the picture of Local Group satellite galaxies has changed drastically in the past few years.  The number of known Milky Way satellites has more than doubled, and their kinematic properties have been carefully characterized \cite{belokurov2007,strigari2008}.  Critically, all known Milky Way satellites have high dark-matter density; the mass within 300 pc of the centers of these satellites is M$_{300} \gtrsim 5\times 10^6 \Modot$ regardless of luminosity.  Any explanation for Milky Way satellite population must correctly reproduce the central density of the satellites as well as the luminosity function and the number of satellites. 

\begin{figure}
  \begin{center}
  \includegraphics[width=0.45\textwidth,angle=270]{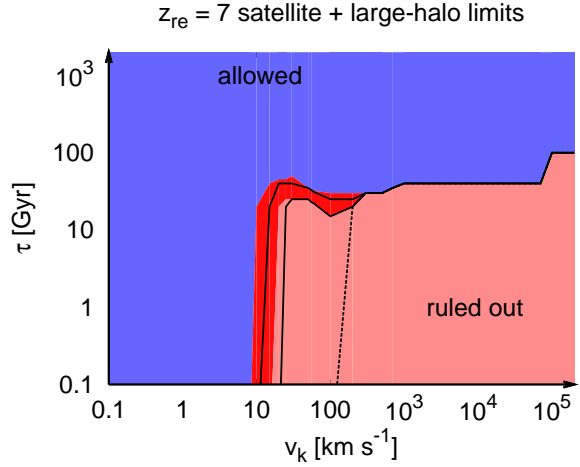}
  \end{center}
  \caption{\label{fig:exclusion}Exclusion limits in the $\tau-\vke$ parameter space.  See text for details.}
\end{figure}

In Ref. \cite{peter2010c}, we used a hybrid technique to construct realizations of Milky Way satellite populations as a function of decay parameters.  We used analytic merger trees for a high-normalization $\Lambda$CDM cosmology in order to find subhalo populations for Milky Way-mass halos.  We used the simulations described in Sec. \ref{sec:sim} to map the properties of the $\Lambda$CDM halos and subhalos to those at $z=0$ assuming specific \vk~and $\tau$.  Finally, we populated subhalos with stars according to a crude star-formation prescription tied to the redshift of reionization $\zre$, for which we take a fiducial value $\zre = 7$.  This star-formation model is meant to illustrate how decay constraints may depend on the as-yet poorly-understood star-formation histories of the satellite galaxies.

We use the prescriptions of Ref. \cite{koposov2008} to estimate the minimum number of satellites with properties similar to the known population likely to be in the Milky Way halo based on the characteristics of the known satellites, taking into account the sky-coverage and completeness of existing searches for Milky Way satellites and the uncertainty in the Milky Way virial mass.  We consider a decay model to be allowed if at least a few of our satellite-population realizations produce at least the minimum number of satellites we expect are in the Milky Way halo.

Our constraints are summarized in Fig. \ref{fig:exclusion}, which shows the allowed and excluded regions of $\vke-\tau$ parameter space assuming that the $X$ particle behaves like cold dark matter.  The pale red region to the right of the dashed curve and under the solid curve is excluded by the galaxy-cluster mass function and the mass-concentration relation.  Local Group satellites provide weaker constraints in this region of parameter space.  The pale red region to the left of the dashed line is that excluded by Milky Way satellites assuming that any subhalo of mass $> 10^7\Modot$ at its accretion time has not been merged with the main halo by dynamical friction; this is an extremely conservative constraint.  The red region is the additional part of parameter space excluded by the Milky Way satellites if dynamical friction in the decaying-dark-matter cosmology is well-described by that in $\Lambda$CDM.   The upper and lower solid lines correspond to the constraints if we assume that all subhalos are populated with stars, with the upper line corresponding to limits if dynamical friction operates as in $\Lambda$CDM, and the lower curve if dynamical friction is ineffective.  

The limits are not very sensitive to the star-formation prescription for satellite galaxies, but are sensitive to the evolution of subhalos and satellites after they are accreted onto larger halos.  Better constraints on decay are only possible if this evolution is better understood.

\section{Discussion}\label{sec:discussion}
In this proceeding, we have described the constraints one may infer on decaying dark matter from different types of observations of dark-matter halos.  The main result for decays is Fig. \ref{fig:exclusion}.  The galaxy-cluster mass function and the mass-concentration relation are most constraining for $\vke \gtrsim 200\hbox{ km s}^{-1}$, and the Milky Way satellites are most constraining for $\vke \lesssim 200 \hbox{ km s}^{-1}$, with $\tau \lesssim 30-40$ Gyr excluded for $\vke \gtrsim 20\hbox{ km s}^{-1}$.

There are several interesting ideas that have emerged as we have done this work.  First, constraints should improve greatly in the future with the advent of truly large and deep sky surveys \cite{lsst2009}.  Second, the cluster mass function and other probes of the growth of structure, traditionally used to constrain the normalization of the matter power spectrum and the nature of dark energy, can be used to constrain the nature of dark matter.  However, the signature of decaying dark matter may be degenerate with that of modified gravity.  This suggests that more work should be done to determine how well next-generation surveys can probe deviations of ``$\Lambda$'' and ``CDM'' simultaneously \cite{peter2010a}.  Third, the distribution of central densities of Local Group satellite galaxies should be a sensitive probe of the nature of dark matter--see Ref. \cite{peter2010c} for further discussion.  


\begin{thebibliography}{99}
\bibitem{komatsu2010} E.~Komatsu et al., arXiv:1001.4538
\bibitem{steigman1985} G.~Steigman \& M.~S.~Turner, Nucl. Phys. B \textbf{253}, 375 (1985)

\bibitem{feng2003} J.~L.~Feng, A.~Rajaraman, \& F.~Takayama, Phys. Rev. Lett. \textbf{91}, 011302 (2003)

\bibitem{feng2008} J.~L.~Feng \& J.~Kumar, Phys. Rev. Lett. \textbf{101}, 231301 (2008)

\bibitem{spergel2000} D.~N.~Spergel \& P.~J.~Steinhardt, Phys. Rev. Lett. \textbf{84}, 3760 (2000)

\bibitem{cohen2010} T.~Cohen, D.~J.~Phalen, A.~Pierce, \& K.~Zurek, Phys. Rev. D \textbf{82}, 056001 (2010)

\bibitem{kaplinghat2005} A.~G.~Doroshkevich \& M.~I.~Khlopov, MNRAS \textbf{211}, 277 (1984); M.~Kaplinghat, Phys. Rev. D \textbf{72}, 063510 (2005); F.~Borzumati, T.~Bringmann, \& P.~Ullio, Phys. Rev. D \textbf{77}, 063514 (2008)

\bibitem{bell2010} Z.~Chen \& M.~Kamionkowski, Phys. Rev. D \textbf{70}, 043502 (2004); L.~Zhang, X.~Chen, M.~Kamionkowski, Z.-G.~Si, \& Z.~Zheng, Phys. Rev. D \textbf{76}, 061301 (2007); N.~.F.~Bell, A.~J.~Galea, \& K.~Petraki, Phys. Rev. D \textbf{82}, 023514 (2010)

\bibitem{peter2010b} A.~H.~G.~Peter, C.~E.~Moody, \& M.~Kamionkowski, Phys. Rev. D \textbf{81}, 103501 (2010)

\bibitem{peter2010c} A.~H.~G.~Peter \& A.~J.~Benson, arXiv:1009.1912

\bibitem{springel2001a} V.~Springel, N.~Yoshida, \& S.~D.~M.~White, New Astronomy \textbf{6}, 79 (2001); V.~Springel, MNRAS \textbf{364}, 1105 (2005)

\bibitem{navarro1996} J.~.F.~Navarro, C.~S.~Frenk, \& S.~D.~M.~White, ApJ \textbf{462}, 563 (1996); J.~.F.~Navarro, C.~S.~Frenk, \& S.~D.~M.~White, ApJ \textbf{490}, 493 (1997)

\bibitem{bryan1998} G.~L.~Bryan \& M.~L.~Norman, ApJ \textbf{495}, 80 (1998)

\bibitem{diemand2009} J.~Diemand \& B.~Moore, arXiv:0906.4340

\bibitem{albrecht2006} A.~Albrecht et al., \emph{Report of the Dark Energy Task Force}, arXiv:astro-ph/0609591

\bibitem{vikhlinin2009} A.~Vikhlinin et al., ApJ \textbf{692}, 1060 (2009); A.~Mantz, S.~W.~Allen, D.~Rapetti, \& H.~Ebeling, MNRAS \textbf{406}, 1759 (2010); E.~Rozo et al., ApJ \textbf{708}, 645 (2010)

\bibitem{peter2010a} A.~H.~G.~Peter, Phys. Rev. D \textbf{81}, 087301 (2010)

\bibitem{bullock2001} J.~S.~Bullock et al., MNRAS \textbf{321}, 559 (2001); R.~H.~Wechsler et al., ApJ \textbf{652}, 71 (2006)

\bibitem{maccio2008} A.~V.~Macci{\`o}, A.~A.~Dutton, \& F.~van den Bosch, MNRAS \textbf{391}, 1940 (2008)

\bibitem{buote2007} D.~A.~Buote et al., ApJ \textbf{644}, 123 (2007); J.~M.~Comerford \& P.~Natarajan, MNRAS \textbf{379}, 190 (2007)

\bibitem{mandelbaum2008} R.~Mandelbaum, U.~Seljak, \& C.~M.~Hirata, JCAP 08 (2008) 006

\bibitem{moore1999} A.~Klypin, A.~V.~Kravtsov, O.~Valenzuela, \& F.~Prada, ApJ \textbf{522}, 82 (1999); B.~Moore et al., ApJ \textbf{524}, L19 (1999)

\bibitem{belokurov2007} V.~Belokurov et al., ApJ \textbf{654}, 897 (2007)

\bibitem{strigari2008} L.~E.~Strigari et al., Nature \textbf{454}, 1096 (2008); M.~G.~Walker et al., Apj \textbf{704}, 1274 (2009) 

\bibitem{koposov2008} S.~Koposov et al., ApJ \textbf{686}, 279 (2008); E.~J.~Tollerud, J.~S.~Bullock, L.~E.~Strigari, \& B.~Willman, ApJ \textbf{688}, 277 (2008)

\bibitem{lsst2009} J.~Annis et al., arXiv:astro-ph/0510195; LSST Science Collaborations, \emph{LSST Science Book, Version 2.0}, arXiv:0912.0201

\end{thebibliography}

\end{document}